**Adoption, usability and perceived clinical value of a UK AI clinical reference platform: a mixed-methods formative evaluation of real-world usage and a 1,223-respondent user survey.**

Kolawole Tytler* [1,2]

[1]NHS, London, UK
[2]University of Cambridge, Cambridge, UK

Email: kt557@cam.ac.uk

* Corresponding author

ORCID: 0000-0002-4553-6971

## ABSTRACT

**Background**

The exponential growth of biomedical literature and clinical guidelines contributes to information overload for healthcare professionals, exacerbating clinician burnout and impeding evidence-based practice. Traditional resources often fail to deliver rapid, accessible information at the point of care. Retrieval-augmented generation (RAG) architectures using large language models (LLMs) offer potential for efficient, provenance-bound clinical decision support, but require evaluation in real-world settings to ensure safety and utility, particularly in public health informatics contexts like UK healthcare.

**Objective**

To describe the design of iatroX, a UK-centred RAG-based clinical reference platform, and report early real-world adoption, usability, and perceived clinical utility from a formative implementation evaluation.

**Methods**

This mixed-methods formative evaluation combined (i) a retrospective observational analysis of platform usage data across web, iOS, and Android applications over 16 weeks (8 April–31 July 2025) and (ii) a cross-sectional, in-product intercept survey.

Usage metrics (unique users, engagement events, clinical queries) were sourced from Google Analytics 4, Apple App Store Connect, and Google Play Console, with bot filtering applied. A client-side script randomized single-item survey prompts to ~10% of web sessions from a predefined battery assessing usefulness, reliability, and adoption intent; responses were anonymous ("Yes/No/Don't Know"). Proportions were reported with Wilson 95% confidence intervals; qualitative comments underwent thematic content analysis. Reporting adhered to STROBE and CHERRIES guidelines. This was classified as a UK service evaluation, requiring no ethics review.

**Results**

The platform achieved rapid adoption with 19,269 unique web users, 202,660 engagement events (~10.5 per active user), and ~40,000 clinical queries across platforms. Mobile downloads included 1,960 iOS and steady Android growth (peak >750 daily active users). The survey elicited 1,223 item-level responses: perceived usefulness 86.2% (95% CI 74.8–93.9%; 50/58); time saved 60.9% (95% CI 38.5–80.3%; 14/23); would use again 93.3% (95% CI 68.1–99.8%; 14/15); willingness to recommend 88.4% (95% CI 75.1–95.9%; 38/43); perceived accuracy 75.0% (95% CI 58.8–87.3%; 30/40); perceived reliability 79.4% (95% CI 62.1–91.3%; 27/34). Themes included speed, guideline-grounded responses, and UK specificity.

**Conclusions**

iatroX demonstrated strong early uptake and positive user perceptions among UK clinicians, highlighting RAG's potential to mitigate information overload in public health informatics. Limitations include small per-item sample sizes and early-adopter bias; future work should include objective accuracy audits and prospective impact studies on workflows and care quality.

**Introduction**

The exponential growth of biomedical literature and clinical guidelines presents a formidable challenge for healthcare professionals striving to practice evidence-based medicine. The volume of medical knowledge is estimated to double at an accelerating rate [1], making it impossible for clinicians to assimilate all relevant information. This phenomenon, termed information overload, requires clinicians to synthesize vast amounts of data, inclusive of national guidelines, local protocols, and pharmacological information, often under significant time constraints at the point of care [2,3]. The cognitive burden associated with navigating this complex information landscape is a significant contributor to clinician burnout and can impede the consistent application of best practices [4].

Traditional digital resources, ranging from PDF documents on institutional intranets to established clinical decision support systems (CDSS) like UpToDate, DynaMed, or BMJ Best Practice, while authoritative, often require time-consuming manual searches and navigation. Studies have shown that clinicians frequently abandon searches if the information is not rapidly accessible, potentially delaying evidence-based care [5,6]. The format of these resources is often not optimized for the rapid, query-based nature of frontline clinical practice.

The emergence of Large Language Models (LLMs) has catalyzed interest in novel approaches to clinical information retrieval and synthesis. LLMs demonstrate remarkable capabilities in natural language understanding and generation, offering the potential to provide instant, conversational answers to complex clinical queries [7]. However, the direct application of general-purpose LLMs (e.g., ChatGPT, Google Gemini) in clinical settings poses significant risks. These models are prone to "hallucination", namely generating plausible-sounding but factually incorrect or outdated information, and often lack transparent provenance for their outputs [8,9]. Furthermore, general LLMs may provide advice that contradicts specific, localized clinical guidelines crucial for safe practice in specific jurisdictions, such as the United Kingdom.

To harness the benefits of LLMs while mitigating these risks, the machine learning field has advanced the Retrieval-Augmented Generation (RAG) architecture [10]. RAG systems enhance LLM performance by grounding the generation process in a specific, external knowledge base. Instead of relying solely on the knowledge internalized during the LLM's training, a RAG system first retrieves relevant documents from a trusted corpus and then uses the LLM to synthesize an answer based on that retrieved information. This approach significantly improves factual accuracy, ensures currency, and provides clear provenance [11].

This paper analyses iatroX, a novel clinical decision support platform utilizing an algorithmic RAG architecture tailored for UK healthcare professionals. iatroX is designed to provide rapid, reliable, and contextually appropriate answers by ensuring all outputs are synthesized directly from a curated, continuously updated knowledge base of authoritative clinical guidelines accepted in UK practice. The platform incorporates a proprietary algorithmic search engine and safety mechanisms to manage uncertainty. By streamlining access to these guidelines, iatroX supports public health informatics goals, such as equitable dissemination of evidence-based resources and reduction of disparities in care delivery across populations.

The objective of this study is to describe the methodology behind the platform and to conduct a mixed-methods formative evaluation of its initial real-world adoption, user engagement, and perceived clinical utility through a large-scale analysis of platform analytics and an in-product user survey.

**Methods**

**Study Design and Setting**

This study employed a mixed-methods formative evaluation design, combining a retrospective observational analysis of real-world usage data with a cross-sectional analysis of in-product user feedback. The study focused on iatroX, a generative AI platform providing clinical decision support for UK healthcare professionals, delivered via a web application and mobile applications (iOS and Android). By facilitating rapid access to evidence-based guidelines, iatroX contributes to public health informatics by promoting consistent application of population-level standards and reducing disparities in care delivery.

Data were collected during a 16-week observational window from 8th April 2025 to 31st July 2025. The study reporting adheres to the Strengthening the Reporting of Observational Studies in Epidemiology (STROBE) statement [12] and the Checklist for Reporting Results of Internet E-Surveys (CHERRIES) [13].

**The iatroX System**

**System Architecture and Regulatory Status**

iatroX is built on a decoupled architecture comprising a Next.js web frontend, a React Native mobile application, and a central Node.js/Express backend API. The primary data store is MongoDB. The system's core functionality relies on a proprietary Retrieval-Augmented Generation (RAG) pipeline.

The platform is registered with the UK Medicines and Healthcare products Regulatory Agency (MHRA) as a Class I Medical Device (Reference: 2025042201417535), adhering to structured quality assurance and software lifecycle processes aligned with standards such as IEC 62304 (Software Lifecycle Processes).

**Retrieval-Augmented Generation (RAG) Pipeline**

The RAG pipeline is engineered to ensure responses are strictly grounded in a verified knowledge base, mitigating the risk of LLM hallucination (Figure 1).

**1. Knowledge Corpus:** The retrieval corpus is constructed from publicly available and accessible UK clinical guidelines and resources from authoritative bodies (e.g., endorsed by NICE, SIGN, Royal Colleges). Ad-hoc scripts monitor these sources daily for updates to ensure content currency.

**2. Ingestion and Indexing:** Documents undergo a proprietary cleaning pipeline and are segmented into semantically coherent text chunks (average 500 tokens, maximum ~3000 tokens). Optimization of this chunking is critical for retrieval relevance [14]. These chunks are converted into numerical vectors using advanced embedding models (GPT embedding family) and indexed in a vector database (Pinecone) for efficient semantic search.

**3. Retrieval and Confidence Scoring:** When a user submits a query, a proprietary algorithmic search engine executes a semantic search, retrieving up to 15 of the most relevant text passages. The algorithm assesses the semantic similarity between the query and the retrieved passages to determine a confidence score.

**4. Dynamic Scope Extension:** If the initial retrieval from the core guidelines yields low confidence scores (indicating potential gaps or ambiguity), the system dynamically extends its scope by triangulating findings with a search of publicly available peer-reviewed research. This secondary search algorithmically ranks evidence by hierarchical strength (e.g., prioritizing meta-analyses over case studies).

**5. Safety Threshold (Refusal to Answer):** If the confidence score remains below a predefined safety threshold after scope extension, the system declines to answer the query. This mechanism is a critical safety feature to prevent the dissemination of information not robustly supported by the evidence base.

**6. Synthesis:** If the confidence threshold is met, the retrieved passages and the original query are passed as context to the generator model. The platform utilizes a cascade of LLMs, including models from the Google Gemini and OpenAI families, alongside a proprietary, post-trained model ("Thea") optimized for clinical synthesis within the iatroX workflow.

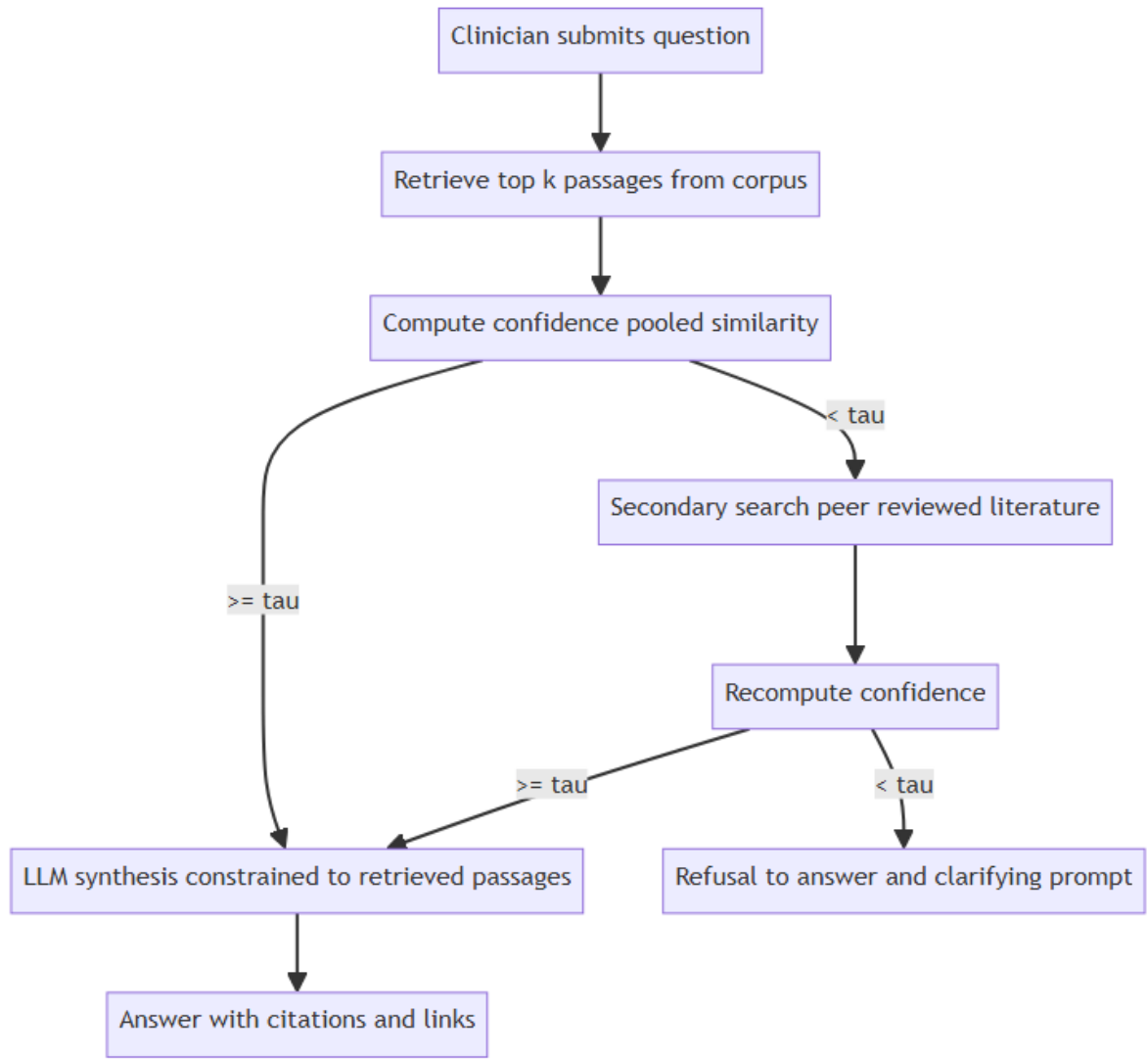

*Figure 1: Safety-aware RAG decision flow. The pipeline retrieves from UK guidelines, extends scope if needed, applies confidence scoring, refuses low-confidence queries, and synthesizes grounded responses with provenance citations (median latency: 12 seconds).*

**Data Collection**

*Usage Analytics*

Platform usage data was collected using Google Analytics 4 (GA4) for the web application, Google Play Console, and Apple App Store Connect for mobile applications. Standard bot filtering was applied. Key metrics included unique users, engagement events (total interactions), active users (DAU/WAU/MAU), and total clinical queries submitted to the RAG pipeline.

*In-Product Intercept Survey*

To capture in-context user perceptions and minimize self-selection bias [15], a systematic random intercept survey was deployed within the web application.

*Sampling and Administration:* A client-side script presented survey prompts to a random 10% of web user sessions. To maximize the diversity of feedback and prevent survey fatigue, users were presented with randomized single-item questions from a larger question battery throughout their session. A browser cookie prevented the same user from being shown the same prompt repeatedly.

*Survey Instrument:* The survey comprised single-item, closed-ended questions assessing domains including perceived usefulness, clinical reliability, system performance, and intent to adopt (Table 2). Responses were recorded as "Yes," "No," or "Don't Know."

*Qualitative Feedback*

Unsolicited qualitative feedback was collected from online blogs, professional social media forums (e.g., Facebook doctor groups), and direct emails to the developer.

**Outcomes and Analysis**

The primary outcomes were the proportions of positive responses ("Yes") to core survey items. Secondary outcomes included descriptive usage metrics (adoption rates, engagement frequency).

Descriptive statistics were used for usage analytics. For survey data, proportions were calculated based on the number of responses received for each specific question. Corresponding 95% Wilson score confidence intervals are reported to estimate the precision of these proportions [16]. Qualitative feedback underwent a rapid thematic content analysis [17] to identify recurrent themes regarding the platform's impact and user trust.

**Ethical Considerations and Governance**

This work was conducted as a service evaluation to assess the performance of a live digital health tool and guide quality improvement. An assessment using the UK Health Research Authority (HRA) decision tool confirmed the project's status as a service evaluation, which does not require Research Ethics Committee (REC) review [18]. The lawful basis for processing anonymized user data was legitimate interests.

**Results**

**Platform Adoption and Engagement**

During the 16-week observational period (8th April 2025 to 31st July 2025), the iatroX platform demonstrated rapid and substantial user adoption (Table 1). The web application attracted 19,269 unique users, generating 202,660 engagement events, indicating a high average interaction rate of 10.5 events per active user. Approximately 40,000 clinical questions were asked across all platforms during the study period.

The user base was predominantly located in the United Kingdom. Traffic analysis revealed that "Direct" traffic was the largest source (9.1k sessions), indicating strong brand recall and habitual return usage. Significant traffic also originated from professional social media groups (e.g., Facebook, 7.7k sessions) and organic search (Google, 4.7k sessions), supporting the observation of organic, word-of-mouth adoption within clinical communities.

Mobile application adoption was robust. The iOS application was downloaded 1,960 times. The Android application demonstrated consistent growth, reaching a peak of over 750 daily active users by late July 2025.

The platform employs a usage-based registration model, allowing guest users three free queries per week before requiring registration. A total of 1,997 users registered during the study period. This represents a visitor-to-registered-user conversion rate of 10.4%, indicating that registration occurred after users had established the utility of the tool (Product-Qualified Users).

**Table 1: Summary of iatroX Platform Usage Metrics (8th April 2025 – 31st July 2025)**

| Metric | Value |
| --- | --- |
| **Web Platform** | |
| Unique Users | 19,269 |
| Total Engagement Events | 202,660 |
| Average Events per User | 10.5 |
| **Mobile Platform** | |
| iOS Downloads | 1,960 |
| Android Peak Daily Active Users | >750 |
| **Cross-Platform** | |
| Total Registered Users | 1,997 |
| Total Clinical Queries Asked (Approx.) | 40,000 |

**User-Perceived Performance and Utility**

A total of 1,223 responses were obtained through the in-product intercept survey. The responses indicated a strongly positive perception across all domains (Table 2).

*Perceived Usefulness and Efficiency*

Users overwhelmingly found the platform beneficial. When asked, "Do you find iatroX useful?", 86.2% (50/58) responded affirmatively. A majority also reported that the platform saved them time (60.9%, 14/23).

*Adoption Intent*

Users indicated a high likelihood of continued use; when asked if they would use iatroX again, 93.3% (14/15) responded "Yes". The willingness to recommend the platform to colleagues was also high (88.4%, 38/43).

*Perceived Clinical Reliability and Trust*

The platform was perceived as clinically accurate and reliable. For the item, "Do you feel that the information provided by iatroX is accurate?", 75.0% (30/40) responded "Yes". When asked, "Does iatroX look and feel reliable to you?", 79.4% (27/34) responded affirmatively.

*System Performance and Usability*

User perception of the system's technical performance and usability was positive. The majority of users found the speed of responses satisfactory (78.9%, 45/57), and the platform easy to navigate (82.2%, 37/45).

**Table 2: User Responses to Key Intercept Survey Items**

| Domain | Survey Item | Yes (%) | 95% CI (Wilson) | No (%) | Don't Know (%) | Total Responses (N) |
|---|---|---|---|---|---|---|
| Usefulness & Efficiency | Do you find iatroX useful? | 86.20% | 75.1–92.9% | 3.40% | 10.30% | 58 |
| | Did iatroX save you time today? | 60.90% | 40.8–77.8% | 13.00% | 26.10% | 23 |
| Adoption Intent | Would you use iatroX again? | 93.30% | 69.9–99.0% | 0.00% | 6.70% | 15 |
| | Would you recommend iatroX to a colleague? | 88.40% | 75.5–95.0% | 2.30% | 9.30% | 43 |
| Reliability & Trust | Do you feel that the information provided by iatroX is accurate? | 75.00% | 59.7–86.1% | 2.50% | 22.50% | 40 |
| | Does iatroX look and feel reliable to you? | 79.40% | 63.2–89.9% | 5.90% | 14.70% | 34 |
| Performance & Usability | Is the speed of responses from iatroX satisfactory? | 78.90% | 67.2–87.3% | 5.30% | 15.80% | 57 |
| | Is the platform easy to navigate on your device? | 82.20% | 69.2–90.6% | 4.40% | 13.30% | 45 |

**Qualitative Feedback Themes**

A large volume of unsolicited qualitative feedback was analyzed. Thematic analysis revealed three recurrent themes:

1. **High Clinical Utility and Efficiency:** Users frequently described the platform as "fantastic," "brilliant," and a "huge service." Feedback emphasized the time saved compared to searching traditional guideline repositories and the platform's ability to synthesize complex management plans rapidly.
2. **Trust Driven by Governance and Provenance:** The MHRA registration (Class I Medical Device) was frequently cited by users as a key differentiator and trust factor compared to general-purpose AI tools. The clear citation of UK-specific guidelines reinforced confidence in the clinical appropriateness of the answers.
3. **Organic Adoption via Professional Networks:** The qualitative data provided numerous examples of users actively sharing the platform within their closed professional circles (e.g., departmental groups, trainee forums), confirming the viral growth observed in the usage analytics.

**Discussion**

**Principal Findings**

This mixed-methods evaluation describes the initial real-world implementation and reception of iatroX, a RAG-based AI clinical decision support platform focused on UK guidelines. The results demonstrate rapid and substantial organic user adoption, with over 19,000 unique web users and 40,000 clinical queries within the first 16 weeks. The in-product survey data from 1,223 respondents indicates that this early adopter cohort perceives iatroX as highly useful (86.2%), accurate (75.0%), and efficient (60.9%). These findings suggest that iatroX successfully addresses a significant unmet need for rapid, trustworthy information retrieval among clinicians.

**Context and Comparison with Existing Literature**

The challenges of information retrieval at the point of care are well-documented drivers of cognitive burden and workflow inefficiency ([19], [20]). The rapid uptake of iatroX underscores the demand for tools that can synthesize authoritative guidelines more efficiently than traditional methods.

The differentiating factor in iatroX's positive reception appears to be its specialized implementation of Retrieval-Augmented Generation (RAG). While there is growing interest in using LLMs in medicine, studies consistently highlight clinician skepticism due to concerns about accuracy, lack of provenance, and potential for "hallucination" ([21], [22]). General-purpose tools, while accessible, are not designed for the safety-critical environment of clinical decision-making and lack the necessary grounding in localized guidelines.

RAG architectures are increasingly recognized as the most viable approach for deploying LLMs in high-stakes, knowledge-intensive domains ([23], [24]). By restricting the LLM's generation process to a curated, verified corpus, in this case, UK-accepted clinical guidelines, iatroX directly addresses these primary safety concerns.

The platform's proprietary algorithmic search engine further enhances this approach. The mechanisms for assessing confidence, dynamically expanding scope to peer-reviewed literature when necessary, and, critically, refusing to answer below a defined safety threshold, provide essential layers of safety. The high level of perceived accuracy (75.0%) and reliability (79.4%) suggests that this RAG implementation is effective in building clinician trust.

Furthermore, the qualitative finding that the MHRA registration fostered trust aligns with research indicating that robust governance and regulatory oversight are critical facilitators for the adoption of clinical AI ([25], [26]).

While other tools exist in the clinical reference space, such as UpToDate, BMJ Best Practice, and newer AI entrants like Glass Health (focusing on differential diagnosis) or approaches utilizing advanced models like Med-Gemini [27], iatroX differentiates itself through its specific focus on UK guidelines, its algorithmic RAG implementation with explicit safety thresholds, and its regulatory status.

The organic, word-of-mouth growth observed (high direct traffic, social media referrals, and 88.4% willingness to recommend) is a strong indicator of product-market fit. It suggests the platform's value proposition is compelling enough for clinicians to recommend it proactively to peers, a powerful mechanism for diffusion of innovation in healthcare [28].

**Strengths and Limitations**

The primary strength of this study is its reliance on a large, real-world dataset, analyzing the behavior of over 19,000 users and capturing 1,223 survey interactions. The use of an in-product intercept survey is a methodological strength, as it captures feedback in the immediate context of use and reduces the recall bias and self-selection bias common in retrospective email surveys [13]. In public health informatics, such tools could extend to population-level applications, like guideline dissemination during outbreaks, enhancing system-wide equity in evidence uptake.

However, the study has several limitations. First, the intercept survey was anonymous, preventing the correlation of user feedback with demographic data (e.g., professional role, specialty, grade). Future research should aim to capture this data to understand adoption patterns across different clinical subgroups. Second, the study population consists of self-selected early adopters, potentially biasing the results towards more positive perceptions; the findings may not generalize to the entire UK healthcare workforce.

Third, the survey utilized single-item, non-validated questions. While appropriate for a formative service evaluation, future evaluations should incorporate validated instruments such as the System Usability Scale (SUS) [15] or the Health Information Technology Usability Evaluation Scale (Health-ITUES) [26]. Fourth, due to the randomized presentation of single items from a larger battery, the sample sizes (N) for individual questions are relatively small. This results in wide confidence intervals for some metrics (e.g., time-saving), limiting the precision of these specific estimates despite the large overall number of respondents.

Finally, while users perceived the information as accurate, this study did not objectively measure the clinical correctness of the generated answers against a gold standard, though it is noted that the iatroX's system is designed to extract, rather than generate, information from the available context.

**Implications for Practice and Future Research**

The positive results of this evaluation suggest that RAG-based AI tools can significantly improve the efficiency of clinical information retrieval. By reducing the time clinicians spend searching for guidelines, these tools may help alleviate cognitive burden and potentially improve the consistency of evidence-based practice.

Future research should focus on rigorous evaluation of the platform's impact. This includes objective assessments of answer accuracy and safety through standardized clinical vignettes reviewed by expert panels. Prospective studies, such as time-and-motion studies comparing iatroX to traditional resources (e.g., NICE website search), are needed to quantify the reported time savings. Ultimately, randomized controlled trials will be necessary to assess the impact of the platform on clinical decision-making quality and patient outcomes.

**Conclusion**

iatroX, a novel RAG-based AI platform registered as a Class I Medical Device, has demonstrated rapid organic adoption and a highly positive reception among a large cohort of UK healthcare professionals. By prioritizing a localized knowledge base, implementing sophisticated retrieval mechanisms with confidence scoring and refusal-to-answer capabilities, and adhering to regulatory standards, the platform has successfully fostered clinician trust. This real-world evaluation provides strong evidence that well-designed and safely implemented RAG systems can meet the critical need for rapid, reliable information synthesis at the point of care, signaling a promising direction for the future of clinical decision support.

**Acknowledgements**

I thank iatroX users who provided in-product feedback during the evaluation period.

**Funding**

No external funding. Platform operations were funded by iatroX.

**Competing interests**

K.T. is the founder and lead developer of *iatroX* and holds equity in the operating company. Usage analytics were obtained from routine telemetry (GA4/App Store Connect/Google Play Console). The survey instrument was pre-specified; anonymised aggregate data and analysis scripts are made available upon reasonable request. No other relationships or activities that could appear to have influenced the submitted work.

**Patient and public involvement**

Patients/the public were not involved in the design, conduct, reporting, or dissemination plans of this research.

**Ethics approval**

This activity was classified using the UK HRA decision tool as a service evaluation; Research Ethics Committee (REC/IRB) review was not required under UK policy. Participation in the in-product survey was voluntary and anonymous; the intercept could be dismissed without responding. Proceeding to the first question constituted implied consent. No personal identifiers or clinical data were collected, and respondents could stop at any time without consequence.

**Consent to participate**

Implied consent was obtained: completion of the voluntary, anonymous survey indicated consent to participate. No personal identifiers were collected.

**Human Ethics and Consent to Participate declarations**

Not applicable (UK service evaluation; REC/IRB review not required; implied consent obtained as described).

**Consent for publication**

Not applicable (no individual person's data in any form).

**Data availability statement**

Anonymised aggregate data, the full survey instrument, and analysis scripts are made available upon reasonable request.